\documentclass[aps,pra,showpacs,floatfix,groupaddress,nofootinbib,twocolumn]{revtex4-1}

\usepackage{graphicx}
\usepackage{amsmath}
\usepackage{epsfig}
\usepackage{amssymb}
\usepackage{mathrsfs}
\usepackage{hyperref}
\hypersetup{colorlinks=true,linkcolor=blue,citecolor=blue,urlcolor=blue}

\begin{document}

\title{Parametric separation of symmetric pure quantum states}

\author{M. A. Sol\'is-Prosser}\email{msolisp@udec.cl} 
\author{A. Delgado}
\affiliation{Center for Optics and Photonics, Universidad de Concepci\'on, Casilla 4016, Concepci\'on, Chile}
\affiliation{MSI-Nucleus on Advanced Optics, Universidad de Concepci\'on, Casilla 160-C, Concepci\'on, Chile}
\affiliation{Departamento de F\'isica, Universidad de Concepci\'on, Casilla 160-C, Concepci\'on, Chile}

\author{O. Jim\'{e}nez}
\affiliation{Departamento de F\'isica, Facultad de Ciencias B\'asicas, Universidad de Antofagasta, Casilla 170, Antofagasta, Chile}

\author{L. Neves}
\affiliation{Departamento de F\'isica, Universidade Federal de Minas Gerais, Belo Horizonte, MG 30123-970, Brazil}


\begin{abstract}
Quantum state separation is a probabilistic map that transforms a given set of pure states into another set of more distinguishable ones. Here we investigate such a map acting onto uniparametric families of symmetric linearly dependent or independent quantum states. We obtained analytical solutions for the success probability of the maps---which is shown to be optimal---as well as explicit constructions in terms of positive operator valued measures. Our results can be used for state discrimination strategies interpolating continuously between minimum-error and unambiguous (or maximum-confidence) discrimination, which, in turn, have many applications in quantum information protocols. As an example, we show that quantum teleportation through a nonmaximally entangled quantum channel can be accomplished with higher probability than the one provided by unambiguous (or maximum-confidence) discrimination and with higher fidelity than the one achievable by minimum-error discrimination. Finally, an optical network is proposed for implementing parametric state separation.
\end{abstract}

\maketitle

\newcommand{\s}{\mathcal{S}}
\newcommand{\f}{\mathcal{F}}

\section{Introduction\label{sec:intro_symm}}

\par The physical maps on quantum states are trace-preserving completely positive maps~\cite{Chefles00Det,Chefles04}. These can be explicitly designed to accomplish a predefined task while simultaneously obeying a set of constraints. For instance, some maps are designed in such a way that they transform a given set of input states onto a fixed set of output states. Two classic examples of this are probabilistic cloning and state discrimination. In the former, a probabilistic map transforms linearly independent states $|\psi_i\rangle$ into states $|\psi_i\rangle|\psi_i\rangle$~\cite{Chefles98qss}, which contains perfect clones. The mapping is constructed to achieve the highest cloning probability. The discrimination of quantum states is realized by mapping nonorthogonal states $|\psi_i\rangle$ to be discriminated onto a set of more distinguishable states, possibly orthogonal ones such as $\{|i\rangle\}$. In this process two different quantities can be optimized: the success probability of the identification and the error in the identification of the states (or confidence). This leads to several discrimination protocols, such as unambiguous state discrimination~\cite{Ivanovic87,Dieks88,Peres88,Jaeger95,Peres98,Chefles98LI}, maximum confidence~\cite{Croke06} and minimum-error~\cite{Holevo73,HelstromBook}, among others.

\par Here, we investigate probabilistic maps between two sets of pure quantum states. In particular, we focus on maps acting onto uniparametric families of symmetric linearly dependent or independent quantum states. These maps are conclusive: it is possible to known with certainty whether the mapping was successful or not. We obtain analytical solutions for the success probability of the maps as well as explicit constructions in terms of positive operator valued measures (POVMs). We also show that the success probability attained by these mappings is optimal. It has been shown recently~\cite{Nakahira12} that any discrimination strategy interpolating between minimum-error discrimination and unambiguous (or maximum-confidence, in the case of linearly dependent states) discrimination can be interpreted as a minimum-error implementation preceded by a map between sets of quantum states. The uniparametric families of symmetric  states studied here are such that an increase in the value of the parameter leads to an increase in the distinguishability of the states. Consequently, these families of states can be used to interpolate continuously between minimum-error discrimination and unambiguous (or maximum confidence) discrimination. For this reason we term these mappings as {\it parametric state separation}. We show that this class of maps finds application in quantum teleportation~\cite{Bennet93} when it is sufficient to teleport through a poorly entangled quantum channel with a fidelity of transmission higher than a given threshold. By using the map proposed here it is possible to achieve the threshold with a higher probability than the one provided by unambiguous state discrimination (or maximum confidence)~\cite{Roa03,Neves12,SolisProsser13} and with a higher fidelity than the achievable by minimum-error discrimination. Application to other protocols such as entanglement swapping and dense coding are also feasible. Finally, we propose a simple optical network to implement parametric state separation.

\par This paper is organized as follows: Section~\ref{sec:separation} proposes a parametrization for symmetric states intended to interpolate between minimum-error and maximum-confidence (or unambiguous) discrimination, that is, the process of parametric state separation. Section~\ref{SSsection_optimal} analyzes the proposed map and shows a closed-form solution for the optimization of the success probability. Section~\ref{SS_applications} displays applications of parametric state separation, such as assisting quantum state teleportation and entanglement swapping. Section~\ref{SSsec:impl1} proposes an optical network able to perform parametric state separation and Section~\ref{SSsummary} concludes this paper.

\section{Parametric state separation \label{sec:separation}}

\par In this work, we address the topic of theory and applications of maps that allow to interpolate between a probabilistic discrimination strategy, as unambiguous or maximum-confidence discrimination, and a deterministic one (e.g., minimum-error discrimination). Since closed-form solutions do not exist for the vast majority of cases, as it also occurs in quantum state discrimination, we shall inspect the context of equally likely symmetric pure states, which has closed-form solutions for several discrimination strategies~\cite{Ban97,Barnett01sym,Jimenez11,Herzog12MC,Chefles98sym}. For this purposes, let us consider a set of $N$ pure quantum states  ${\{|\psi_{j}\rangle,\,j=0,\dots,N-1\}}$ spanning a $D$-dimensional Hilbert space, with $N \geqslant D$. These states are \emph{symmetric} under the action of a unitary operator $\hat{W}$ if they satisfy 
\begin{align}
|\psi_{j}\rangle &= \hat{W}|\psi_{j-1}\rangle = \hat{W}^{j} |\psi_0\rangle,\\
|\psi_0\rangle &= \hat{W}|\psi_{N-1}\rangle . 
\end{align}
	For every unitary operator $\hat{W}$, there exist a set of phases and eigenvectors $\{(\zeta_k ,|\gamma_k\rangle),~k=0,\dots,N-1\}$ such that 
\begin{equation}
	\hat{W}=\sum_{k=0}^{N-1}e^{i\zeta_k}|\gamma_k\rangle\langle \gamma_k|.
\end{equation} 
The state $|\psi_0\rangle$ is sometimes known as the \emph{fiducial} state of the set. The condition ${\hat{W}|\psi_{N-1}\rangle=|\psi_0\rangle}$ implies that ${\hat{W}^N=\hat{I}}$. For this to be fulfilled, it is necessary that ${N\zeta_k=2\pi n_k,~n_k\in\mathbb{Z}}$. Then, every unitary operator $\hat{W}$ able to generate a set of $N$ symmetric $D$-dimensional states can be written as 
\begin{equation}
\hat{W}=\sum_{k=0}^{N-1}e^{2\pi i n_k/N}|\gamma_k\rangle\langle\gamma_k|.
\end{equation}
As remarkable examples, the eigenstates of the phase gates $\hat{Z}|l\rangle = e^{2\pi i l/N}|l\rangle$ correspond to $|\gamma_k\rangle=|k\rangle$, and the ones of the shift gates $\hat{X}|l\rangle=|l\oplus 1\rangle$ correspond to  the inverse Fourier transform of the computational basis, i.e., $|\gamma_k\rangle = \hat{F}^{-1}|k\rangle$.

\par Now, let us define two sets of $N$ symmetric states spanning a $D$-dimensional Hilbert space, defined by
\begin{align}
	|\alpha_{j}\rangle  =&  \sum_{k=0}^{N-1} a_k e^{i\phi_k}\omega^{jk}|\mathbf{f}_k\rangle,\label{eq:alpha}\\
	|\beta_{j}\rangle   =&  \sum_{k=0}^{N-1} b_k e^{i\varphi_k}\omega^{jk}|\mathbf{g}_k\rangle, \label{eq:beta}
\end{align}
where $\omega=\exp(2\pi i/N)$ is a $N$th complex root of the unity, $a_k$ and $b_k$ are real non-negative coefficients where $N-D$ of them adopt null values, $\phi_k$ and $\varphi_k$ are real numbers representing phases, $a_ke^{i\phi_k}$ are the coefficients of the fiducial state of the first set, $b_ke^{i\varphi_k}$ are those ones of the second set, $|\mathbf{f}_k\rangle$ and $|\mathbf{g}_k\rangle$ are orthonormal eigenbasis for the operators that generate each set. The first set is symmetric under ${\hat{Z}_\alpha=\sum_{k=0}^{N-1} \omega^k |\mathbf{f}_k\rangle\langle \mathbf{f}_k|}$, and the second one is symmetric under ${\hat{Z}_\beta=\sum_{k=0}^{N-1} \omega^k |\mathbf{g}_k\rangle\langle \mathbf{g}_k|}$. Although the operators $\hat{Z}_\alpha$ and $\hat{Z}_\beta$ were written as phase operators, this representation holds only when they are written in their respective eigenbases. 

\par The general task of finding an operation $\Lambda$ such that $\Lambda(|\alpha_{j}\rangle)\propto |\beta_{j}\rangle$ has been studied. Currently, results about existence and feasibility of these physical operations are known~\cite{Chefles00Det,Chefles02gamma,Chefles04,Feng05,Zhou07}. However, closed-form solutions are known for sets of two states only~\cite{Chefles98qss,Roa10,TorresRuiz09}. Additionally, it was proven that optimal maps between sets of symmetric states can be found by solving a linear programming problem~\cite{Dunjko12}. This optimization problem can always be numerically solved. Nevertheless, closed-form solutions and details about its implementation, i.e., the corresponding Kraus operators, must be studied for every particular case.

\par Suppose we have two sets of $N$ pure states defined by
\begin{align}
	|\alpha_{j}\rangle  =&  \sum_{k=0}^{N-1} a_k e^{i\phi_k}\omega^{jk}|k\rangle, \label{eq:alphaj}\\
	|u_{j}\rangle  =&  \frac{1}{\sqrt{{ D }}}\sum_{k=0}^{N-1} y_k \omega^{jk}|k\rangle,\label{eq:uj}
\end{align}
where $y_k$ is a binary parameter such that
\begin{align}
	y_k=\begin{cases}
			1 &{\rm if} ~~ a_k\neq 0\\
			0 &{\rm if} ~~ a_k = 0
	\end{cases},\label{defyk}
\end{align}
and ${ D }=\sum_{k=0}^{N-1}y_k$, where $D$ is the dimension of the Hilbert space in which states $|\alpha_j\rangle$ lie and is also the number of non-vanishing coefficients $a_k$. The states $|u_{j}\rangle$ constitute a set of $D$ orthonormal states when ${N=D}$. Otherwise, they compose a set with maximal distinguishability within the ${ D }$-dimensional subspace where they belong.  Optimal operations transforming states $|\alpha_{j}\rangle$ onto states $|u_{j}\rangle$ have been studied~\cite{Chefles98sym,Jimenez11,Herzog12MC}. For instance, the problem of discriminating within an ensemble of equally likely symmetric states was studied via Unambiguous discrimination~\cite{Ivanovic87,Dieks88,Peres88,Jaeger95,Peres98} for linearly independent states~\cite{Chefles98LI} and Maximum Confidence~\cite{Croke06} for linearly dependent states. 

Here, we consider maps from the set of states $|\alpha_{j}\rangle$ to the set of states $|\beta_{j}(\xi)\rangle$ defined as
\begin{align}
	|\beta_{j}(\xi)\rangle &= \sum_{k=0}^{N-1} b_k(\xi)e^{i\varphi_k(\xi)}\omega^{jk}|k\rangle,\label{eq:betasuc}
\end{align}
where
\begin{align}
	b_k(\xi)^2  &=  \left[a_k^2 + \left(\frac{1}{{ D }} - a_k^2\right)\xi\right]y_k \nonumber\\
	 &=  (1-\xi)a_k^2 + \frac{y_k\xi}{{ D }},\label{eq:bkcoef}
\end{align}
are the coefficients of the fiducial output states. As ${|\beta_{j}(0)\rangle=|\alpha_{j}\rangle}$ and ${|\beta_{j}(1)\rangle=|u_{j}\rangle}$, then $\xi$ is a parameter that indicates a point between the sets $|\alpha_{j}\rangle$ and $|u_j\rangle$ in which the states $|\beta_{j}(\xi)\rangle$ are lying on, with ${0\leq \xi \leq 1}$. The phase $\varphi(\xi)$ must equal $\phi_k$ for $\xi=0$, and $0$ for $\xi=1$. A simple linear formula, which will be used here\footnote{Other choices for $\varphi_k(\xi)$ are possible. For instance, $\varphi_k(\xi)=-\phi_k$, which immediately eliminates the phases of the fiducial state no matter the value of $\xi$.}, is $\varphi_k(\xi) = (1-\xi)\phi_k$. Note, from Eq.~(\ref{eq:bkcoef}), that we have considered $a_ky_k=a_k$, as can be inferred from the definition of Eq.~(\ref{defyk}). The way these numbers were defined is based upon the parametric definition of a line segment between two points, in which the endpoints are $a_k^2$ and $1/{ D }$. This definition for $b_k(\xi)$ preserves the norm of the states and the ordering of the coefficients, i.e., ${a_k\geqslant a_l}$ implies that  ${b_k(\xi)\geqslant b_l(\xi)}$ and ${a_m=0}$ implies that ${b_m(\xi)=0}$, for every value of $\xi$ in the domain.

\begin{figure}[!t]
	\centering
	\includegraphics[width=0.98\columnwidth]{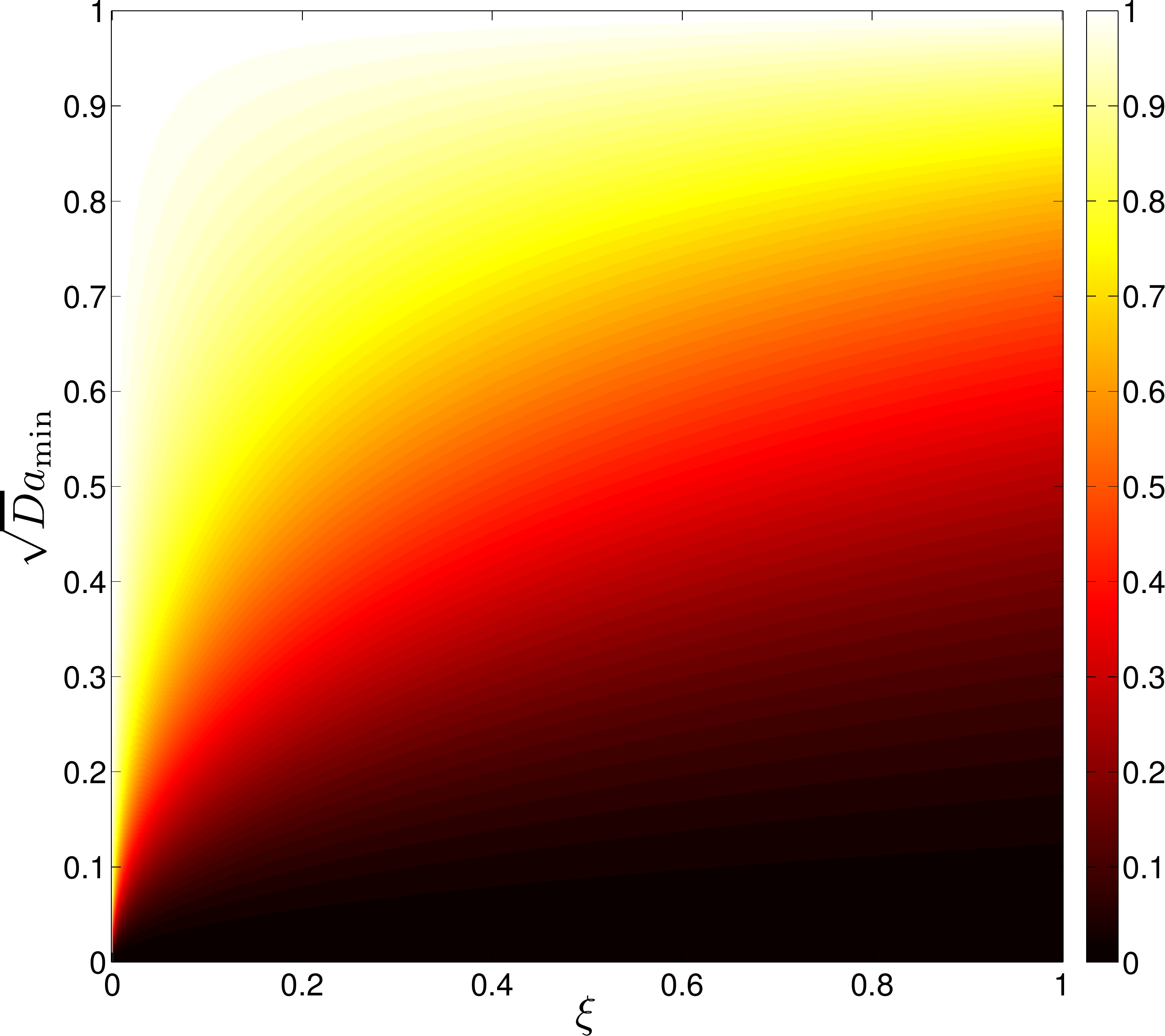}
	\caption{(Color online) Success probability $p_\s(\xi)$ of Eq.~(\ref{eq:psxi}) for parametric state separation as a function of the minimum coefficient $a_{\min}$ (scaled by the number of non-vanishing fiducial coefficients ${ D }$) and the $\xi$ parameter that defines the target set of states.\label{prob_sep}}
\end{figure}

\begin{figure*}[!t]
	\centering
	\includegraphics[width=0.98\textwidth]{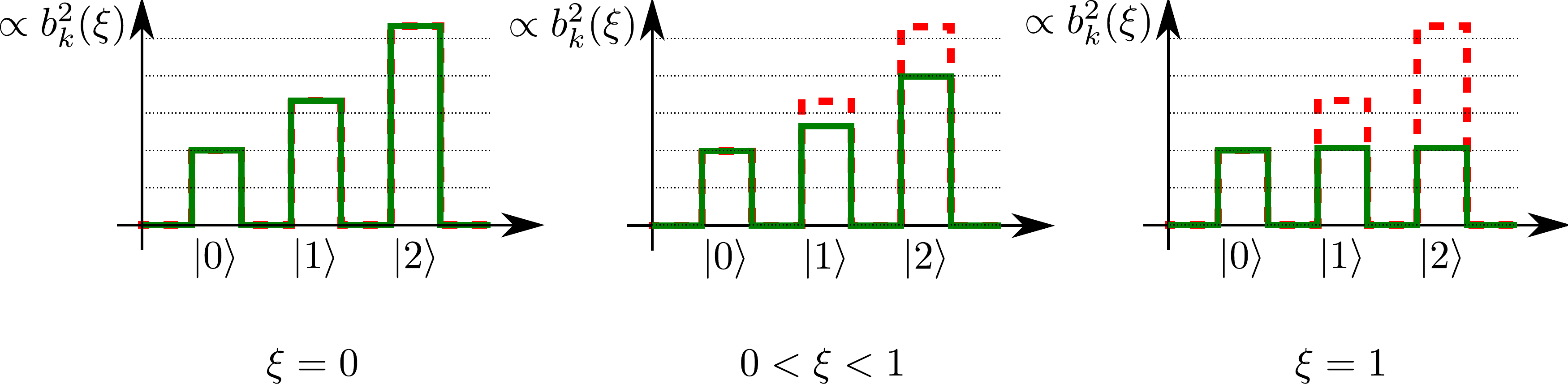}
	\caption{(Color online) Simplified depiction of the coefficients during parametric state separation. The dashed line represents the initial coefficients $a_k^2$, which are equal to $b_k^2(0)$. The continuous line depicts the values of $b_k^2(\xi)$ for $\xi>0$, showing the relevant changes on these probabilities.\label{fig:comparison}}
\end{figure*}

\par A procedure that implements the mapping of states $|\alpha_{j}\rangle_s$ onto states $|\beta_{j}(\xi)\rangle_s$ is given by~\cite{He06}
\begin{align}
	\hat{\mathcal{U}}_{sa}(\xi)|\alpha_{j}\rangle_s|\mathscr{A}\rangle_a  =&  \hat{A}_\s(\xi)|\alpha_{j}\rangle_s|0\rangle_a + \hat{A}_\f(\xi)|\alpha_{j}\rangle_s|1\rangle_a, \nonumber \\
	 =&  \sqrt{p_\s(\xi)}|\beta_{j}(\xi)\rangle_s|0\rangle_a\nonumber \\ &+ \sqrt{1-p_\s(\xi)}|\widetilde{\beta}_{j}(\xi)\rangle_s|1\rangle_a,\label{eq:unitary}
\end{align}
where $|0\rangle_a$ and $|1\rangle_a$ are two orthogonal states from a two-dimensional ancillary system, $\hat{\mathcal{U}}_{sa}(\xi)$ is a unitary operator acting on the bipartite Hilbert space of the system ($s$) and the ancilla ($a$), $|\mathscr{A}\rangle_a$ is the initial state of the ancilla and $p_\s(\xi)$ is the probability of achieving the desired mapping. A measurement on the ancilla performed afterwards can announce whether the map was successful. In this case, it has been assumed the success probability to be the same for all $|\alpha_{j}\rangle$ states. The \emph{uniformization lemma} of Ref.~\cite{Dunjko12} supports this assumption. The Kraus operators $\hat{A}_\s(\xi)$ and $\hat{A}_\f(\xi)$ represent the process of mapping successfully and unsuccessfully, respectively. Since $\hat{\mathcal{U}}(\xi)$ is a unitary operation, these Kraus operators must satisfy ${\hat{A}_\s^\dagger(\xi)\hat{A}_{\s}^{\vphantom{\dagger}}(\xi)+\hat{A}_\f^\dagger(\xi)\hat{A}_{\f}^{\vphantom{\dagger}}(\xi)=\hat{I}}$, where $\hat{I}$ is the identity operator acting on the Hilbert space of the symmetric states. The Kraus operator related to successful events can be written as
\begin{align}
\hat{A}_{\s}^{\vphantom{\dagger}}(\xi) = & \sqrt{p_\s}\sum_{k=0}^{N-1} \frac{b_k(\xi)}{a_k} e^{-i\xi\phi_k } |k\rangle\langle k|,
\end{align}
where only the terms with non-vanishing coefficients of the fiducial state contribute. The probability of success can be maximized under the constraint that ${\hat{A}_\f^\dagger(\xi)\hat{A}_\f(\xi)=\hat{I}-\hat{A}_\s^\dagger(\xi)\hat{A}_\s(\xi)\geqslant0.}$ This constraint ensures the non-negativity of the probability of failure given by ${p_\f=1-p_\s.}$  As a result, 
\begin{align}
	p_\s(\xi)  =&  \min_{\{k\}}\left[\frac{a_k^2}{b_k^2(\xi)}\right] 
	= \frac{1}{(1-\xi)+\frac{\xi}{{ D }a_{\min}^2}}, \label{eq:psxi}
\end{align}
where $a_{\rm min}=\min\{a_j|j=0,\dots,N-1\, \wedge \,a_j\neq 0\}$. A graph of the success probability is shown in Fig.~\ref{prob_sep} in terms of the smallest coefficient of the initial fiducial state and number of non-vanishing coefficients ($y$-axis) and the target family of states determined by $\xi$ ($x$-axis). Additionally, it can be shown that 
\begin{equation}
	b_{\min}^2(\xi)= (1-\xi)a_{\min}^2 + \xi/{ D } \label{eq:bmin}
\end{equation}
is the smallest of the non-vanishing $b_k^2(\xi)$ coefficients. The Kraus operators $\hat{A}_\s(\xi)$ and $\hat{A}_\f(\xi)$ can then be written as
\begin{align}
	\hat{A}_{\s}^{\vphantom{\dagger}}(\xi)= & \sum_{k=0}^{N-1}  \sqrt{\frac{1-\xi+\xi/{ D }a_k^2}{1-\xi+\xi/{ D }a_{\rm min}^2}} y_k  e^{-i\xi\phi_k }|k\rangle\langle k|,\label{eq:As} \\
	\hat{A}_{\f}^{\vphantom{\dagger}}(\xi)= & \sum_{k=0}^{N-1} \Bigg[\sqrt{\frac{\xi}{{ D }}\frac{1/a_{\rm min}^2-1/a_k^2}{1-\xi+\xi/{ D }a_{\rm min}^2}   } e^{-i\xi\phi_k } y_k\nonumber\\& \hspace{3.5cm} + (1-y_k)\Bigg] |k\rangle\langle k| . \label{eq:Af}
\end{align}
\par The action of these operators on the states $|\alpha_{j}\rangle$ is given by 
\begin{align}
	\hat{A}_{\s}^{\vphantom{\dagger}}(\xi)|\alpha_{j}\rangle 	= & \sqrt{p_\s(\xi)}|\beta_{j}(\xi)\rangle,\\
	\hat{A}_{\f}^{\vphantom{\dagger}}(\xi)|\alpha_{j}\rangle 	= & \sqrt{1-p_\s(\xi)}|\widetilde{\beta}_{j}(\xi)\rangle, \label{eq:betafail}
\end{align}
where
\begin{align}
|\widetilde{\beta}_{j}(\xi)\rangle = \sum_{k=0}^{N-1} \sqrt{\frac{a_k^2 - a_{\min}^2 y_k}{1-{ D }a_{\min}^2}} \omega^{jk}e^{i(1-\xi)\phi_k } |k\rangle.\label{eq:betatilde}
\end{align}

Recapitulating, once the bipartite unitary operation has been carried out, the user must perform a measurement on the ancilla. If the outcome is $|0\rangle_a$ ($|1\rangle_a$), the state $|\alpha_{j}\rangle$ has been mapped on $|\beta_{j}(\xi)\rangle$ ($|\widetilde{\beta}_{j}(\xi)\rangle$) and the process is considered successful (failed). The states $|\beta_{j}(\xi)\rangle$ and $|\widetilde{\beta}_{j}(\xi)\rangle$ are given by Eqs.~(\ref{eq:betasuc}) and~(\ref{eq:betatilde}), respectively. It is worth mentioning that, although the components $|\langle k|\beta_{j}(\xi)\rangle|=b_k(\xi)$ of the successfully transformed states depend on $\xi$, the coefficients $|\langle k|\widetilde{\beta}_{j}(\xi)\rangle|$ are not functions of $\xi$. Additionally, there always be at least one value of $k$, say $k_0$, such that $a_{k_0}=a_{\min}$ and, correspondingly, $b_{k_0}(\xi)=b_{\min}(\xi)$, making the ${k_0}$th coefficient of $|\widetilde{\beta}_{j}(\xi)\rangle$ vanish in the case of a failed attempt. Then, the smallest coefficient(s) is (are) not transferred to the failure states, which are even less distinguishable among them than the original set $\{|\alpha_{j}\rangle\}$ since it is the same number of vectors distributed over a smaller subspace.

It is useful to interpret these results in terms of amplitude modulation. Figure~\ref{fig:comparison} sketches this topic through a three-dimensional example. This analysis can be applied for every physical system able to encode a set of symmetric states, where the rectangle functions represent the squared amplitude of, for instance, the energy levels of an atom, the paths of a multiport interferometer, an array of slits, among others examples. It can be seen that the smallest coefficient keeps unaltered and the larger coefficients are attenuated till they reach the value of the smaller one according the parameter $\xi$ increases. The closer is $\xi$ to 1, the lesser the difference among the coefficients. As it can be observed from Fig.~\ref{fig:comparison}, the probabilities for each basis component is reduced according $\xi$ increases. The difference between the original probabilities of the initial states and the final ones, for each $\xi>0$, will construct the unsuccessful case and, consequently, the coefficients of failure states $|\widetilde{\beta}_j(\xi)\rangle$. This is consistent with Eq.~(\ref{eq:betatilde}) as they have null probability in the basis vector corresponding to the smallest coefficient.

\subsection{Distinguishability measures}

\par The map of states $|\alpha_j\rangle$ onto $|\beta_j(\xi)\rangle$ defined by Eqs.~(\ref{eq:alphaj}) and~(\ref{eq:betasuc}), respectively, is called parametric state separation, with the parametrization defined by Eq.~(\ref{eq:bkcoef}). A natural question that arises here concerns the distinguishability of the resulting states $|\beta_j(\xi)\rangle$. From the geometrical point of view, two vectors have been \emph{separated} if the angle between them at the end is larger than at the beginning. For more states, we must resort to a distinguishability measure in order to show that the parametrization given by Eq.~(\ref{eq:bkcoef}) makes the transformed states more distinguishable than the initial ones and the opposite behavior is not occurring. There exist several quantities to measure distinguishability among the states belonging to a given set. In this work, we will make use of two of them. The first one is the optimal probability of having a successful discrimination via unambiguous discrimination or maximum-confidence---according to their linear dependence---among these states, which we shall define as $\mathcal{D}_1(\xi)$. The second one is the optimal probability of correctly discriminating these states by means of optimal minimum-error discrimination, denoted as $\mathcal{D}_2(\xi)$ throughout this section. Since the states we are dealing with are symmetric, these two quantities are known~\cite{Chefles98sym,Jimenez11,Ban97}. The use of at least two different measures is necessary because there exist sets of states whose ordering, by distinguishability, is different depending on which measure is used~\cite{Chefles02}. For this reason, there is not an absolute distinguishability measure within an ensemble of quantum states and, consequently, the use of only one measure is not reliable.

\begin{figure}[!t]
	\centering
	\hspace{-0.35cm}\includegraphics[width=1.04\columnwidth]{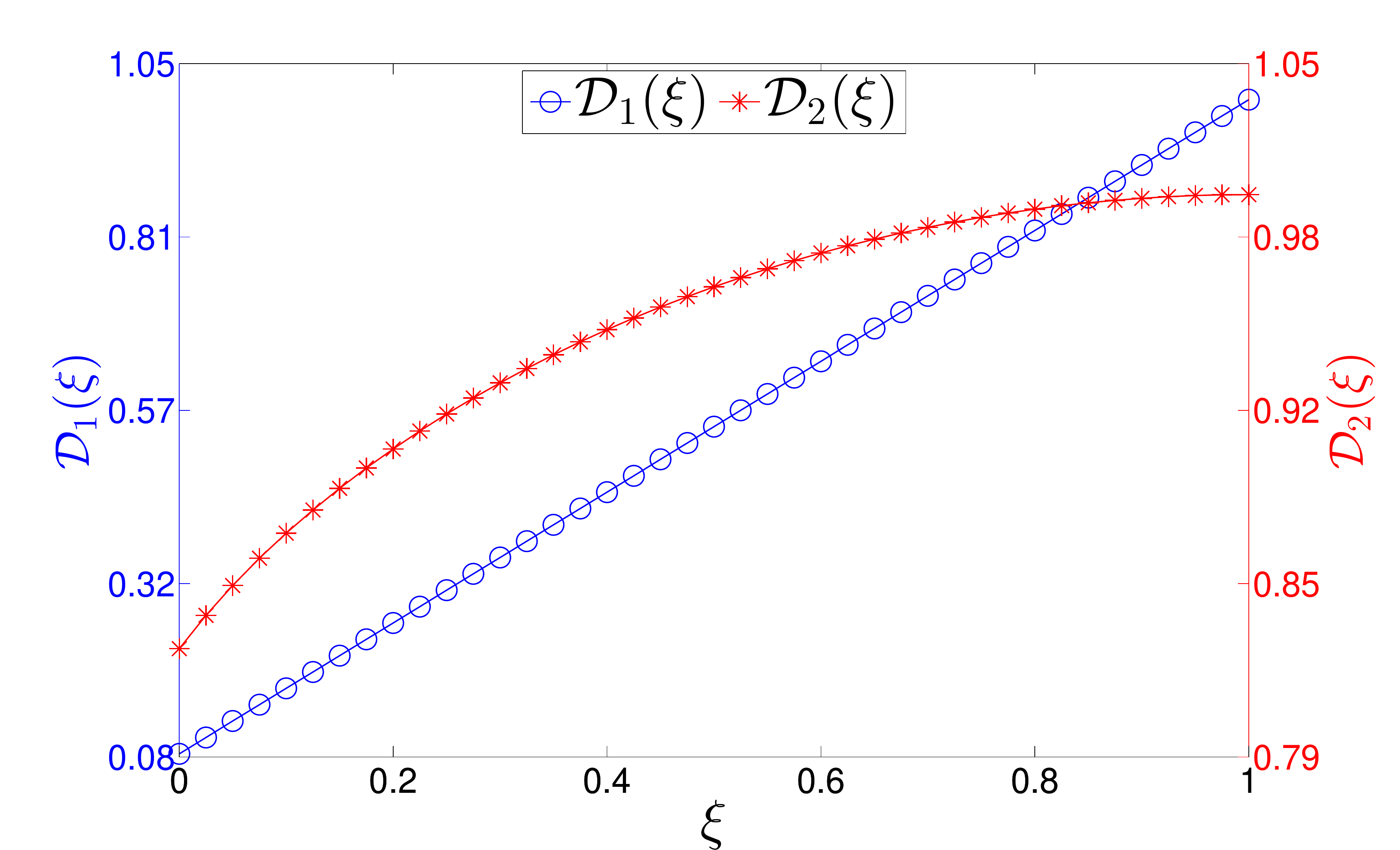}
	\caption{(Color online) Example of distinguishability measures behavior according $\xi$ increases. For this example, ${N=D=5}$, ${a_0=0.6386}$, ${a_1=0.5841}$, ${a_2=0.3817}$, ${a_3=0.1321}$, and ${a_4=0.2964}$. Left vertical axis represents the first distinguishability measure given by Eq.~(\ref{eq:sep1}), which is shown in the graph as circles. Conversely, right vertical axis stands for the second measure defined by Eq.~(\ref{eq:sep2}), represented in the graph as stars. For this particular example, both measures are increasing functions of $\xi$. A proof of this behavior for any set of initial states can be found in the main text. \label{fig:distinguishability}}
\end{figure}

\par Figure~\ref{fig:distinguishability} shows an example of these measures applied on a given fiducial state. As expected, both increase according $\xi$ adopts larger values. But, beyond a single example, a more rigorous proof of the increased distinguishability of states $|\beta_j(\xi)\rangle$ with respect to $|\alpha_j\rangle$  is necessary. According to the first measure and taking Eq.~(\ref{eq:bmin}) into account, we obtain~\cite{Chefles98sym,Jimenez11,Herzog12MC}
\begin{align}
	\mathcal{D}_1(\alpha) =& { D }a_{\min}^2, \\
	\mathcal{D}_1(\beta(\xi)) =&  { D }b_{\min}^2(\xi) \nonumber \\
		=&\mathcal{D}_1(\alpha) + (1-{ D }a_{\min}^2)\xi.\label{eq:sep1}
\end{align}
Since $a_{\min}\leqslant1/\sqrt{{ D }}$, the second term of the RHS of Eq.~(\ref{eq:sep1}) is always positive. Then, for every value of $\xi\in[0,1]$, it occurs that $\mathcal{D}_1(\beta(\xi))\geqslant\mathcal{D}_1(\alpha)$ and the states $|\beta_{j}(\xi)\rangle$ are more distinguishable than the states $|\alpha_{j}\rangle$ according to the first measure function. Moreover, distinguishability increases as $\xi$ grows.

On the other hand, the second measure is the probability of optimal minimum-error discrimination~\cite{Holevo73,HelstromBook}, given by~\cite{Ban97,Barnett01sym}
\begin{align}
	\mathcal{D}_2(\alpha)  =&  \frac{1}{N}\left(\sum_{k=0}^{N-1} a_k\right)^2, \\
	\mathcal{D}_2(\beta(\xi))  =&  \frac{1}{N}\left(\sum_{k=0}^{N-1} b_k(\xi)\right)^2.\label{eq:sep2}
\end{align}
The behavior of the second distinguishability measure can be studied if we calculate its derivatives with respect to $\xi$. For this purposes, we shall define $\mathcal{E}(\xi) = \sqrt{N\mathcal{D}_2(\beta(\xi))}$, whose behavior is simpler to analyze. So, from Eqs.~(\ref{defyk}) and~(\ref{eq:bkcoef}), 
\begin{align}
	\mathcal{E}(\xi) &= \sum_{k=0}^{N-1} b_k(\xi) \nonumber \\
		 & = \sum_{k=0}^{N-1} y_k \sqrt{ (1-\xi)a_k^2 + \xi/{ D }}.
\end{align}
Then, 
\begin{align}
	\frac{{\rm d}}{{\rm d}\xi}\mathcal{E}(\xi) &=  \sum_{k=0}^{N-1} y_k \frac{1/{ D } -a_k^2}{2\sqrt{ (1-\xi)a_k^2 + \xi/{ D }}}, \\
	\frac{{\rm d}^2}{{\rm d}\xi^2}\mathcal{E}(\xi) &= -\sum_{k=0}^{N-1} y_k \frac{(1/{ D } -a_k^2)^2}{4[(1-\xi)a_k^2 + \xi/{ D }]^{3/2}} \leqslant 0,~~\forall~~\xi.
\end{align}
We can see that the graph of $\mathcal{E}(\xi)$ is concave downwards since its second derivative is always negative. Although it is not easy to find critical points from its first derivative, we can observe for $\xi=1$ that 
\begin{align}
	\frac{{\rm d}}{{\rm d}\xi}\mathcal{E}(1) &=  \sum_{k=0}^{N-1} y_k \frac{1/{ D } -a_k^2}{2\sqrt{ (1-1)a_k^2 + 1/{ D }}} \nonumber \\
	&=  \frac{\sqrt{ D }}{2}\left(\frac{1}{{ D }}\sum_{k=0}^{N-1}y_k - \sum_{k=0}^{N-1} a_k^2 \right) = 0.
\end{align}
Then, $\xi=1$ is a critical point. The second derivative in this point is negative, so this is a maximum. This is expected because $\xi=1$ leads to orthogonal states when ${N=D}$. Since $\mathcal{E}(\xi)$ is concave downwards and has its maximum in $\xi=1$, we can ensure that $\mathcal{E}(\xi)$ is monotonically increasing for $0\leqslant \xi\leqslant 1$ and, in consequence, $\mathcal{D}_2 = \mathcal{E}^2(\xi)/N$ is also an increasing function. Therefore, this distinguishability measure increases according $\xi$ does and $\mathcal{D}_2(\beta(\xi))\geqslant \mathcal{D}_2(\beta(0)) = \mathcal{D}_2(\alpha)$. 

As shown in this section, there exist at least two distinguishability measures that consider the states $|\beta_j(\xi)\rangle$ as more distinguishable than the $|\alpha_j\rangle$. Then, the parametrization proposed in Eq.~(\ref{eq:betasuc}) is correct in terms of the initial goal of an intermediate map between the original states $|\alpha_j\rangle$ and the states $|u_j\rangle$ of Eq.~(\ref{eq:uj}). This opens the possibility of considering alternative strategies in state discrimination equivalent to perform a map onto a more distinguishable set followed by minimum error discrimination over the final set~\cite{Nakahira12}.

\section{The optimal transform\label{SSsection_optimal}}
\subsection{Maps between symmetric pure states as a linear programming problem}
In order to map from the symmetric states $|\alpha_{j}\rangle$ to the symmetric ones $|\beta_{j}\rangle$, we can also regard a more general transformation for pure states, given by~\cite{Dunjko12}
\begin{align}
 	\hat{\mathcal{U}}_{sea}|\alpha_{j}\rangle_s|\mathcal{M}\rangle_e|\mathcal{A}\rangle_a\ =& \sqrt{p}|\beta_{j}\rangle_s|\psi_{j}\rangle_e|0\rangle_a \nonumber\\ &+ \sqrt{1-p}|\Phi_{j}\rangle_{se}|1\rangle_a, \label{eq:three_system}
\end{align}
where $p$ is the success probability, $|\psi_{j}\rangle$ are states generated as a by-product of the process and $|\Phi_{j}\rangle_{se}$ are the resulting states when the process fails. This transformation differs from the one of Eq.~(\ref{eq:unitary}) since, in addition to the main system ($s$) and ancilla ($a$), it includes a third system representing the environment ($e$) or a second auxiliary system. These three-system transforms may allow higher success probability than two-system transforms~\cite{Zhou07}. It is noteworthy that transforms like the one of Eq.~(\ref{eq:unitary}) are particular cases of Eq.~(\ref{eq:three_system}) in which the environment state evolves independently of the other two systems. Dunjko and Andersson showed that if such transform exists, then there exist an equivalent one in which the $|\psi_{j}\rangle$ states are also symmetric. This result was shown in the \emph{symmetrization lemma}~\cite{Dunjko12}. Additionally, they proved that the transform mapping $|\alpha_j\rangle$ to $|\beta_j\rangle$ with optimal probability of success can be found by solving a linear problem which, when applied to the states studied throughout this paper, can be summarized as 
\begin{align}
	\max_{\mathbf{x}} & ~~\mathbf{c^\intercal x} = p \nonumber\\
	{\rm subject\,~to: }& \begin{cases} 0\leqslant \mathbf{x}\leqslant 1 \\ \mathbf{Mx\leqslant n}  \end{cases}, \label{linear_problem}
\end{align}
where 
\begin{align}
\mathbf{n}=&\sum_{r=0}^{N-1} a_r^2 \vec{\mathbf{e}}_r,\hspace{1cm}
\mathbf{c}=\sum_{k=0}^{N-1}\vec{\mathbf{e}}_k,\nonumber	\\
\mathbf{M}=&\sum_{r=0}^{N-1}b_r^2\hat{\bf X}^r,\hspace{1cm}
\hat{\bf X}=\sum_{k=0}^{N-1}\vec{\mathbf{e}}_{k+1} \vec{\mathbf{e}}_{k}^\dagger ,
\end{align}
are fixed vectors and matrices, $\{\vec{\mathbf{e}}_j\}_{j=0}^{N-1}$ is the $N$-dimensional computational Euclidean basis, and 
\begin{equation}
\mathbf{x}=\sum_{k=0}^{N-1} p\Psi_{k}^{2} \vec{\mathbf{e}}_k\label{eq:xvector}
\end{equation}
is the unknown $N$-dimensional vector to be found~\cite{Dunjko12}. The parameters $\Psi_s$ are also the coefficients of the state $|\psi_0\rangle$ and they give information about the Kraus operators involved in the transform, as shown in the Appendix. A transform in which all of the $|\psi_{j}\rangle$ states are equal is referred to as a \emph{leakless} transform~\cite{Dunjko12}, whose solution has the form $\mathbf{x}=p\vec{\mathbf{e}}_{k_0}$ for a certain $k_0$. Examples of leakless transforms include the ones of Eq.~(\ref{eq:unitary}). Particularly, Eq.~(\ref{eq:psxi}) is represented by ${\mathbf{x}=p_\s(\xi)\vec{\mathbf{e}}_0}$.

\par Although linear programming problems like the one of Eq.~(\ref{linear_problem}) can always be numerically solved, the search of closed-form solutions for algebraic coefficients and arbitrary dimensions and number of states is not always an easy task to solve. Nevertheless, we were able to proof that the leakless transform of Eq.~(\ref{eq:unitary}) actually represents the \emph{optimal} procedure to carry out parametric state separation, since any physically acceptable solution different from $\mathbf{x}=p_\s(\xi)\vec{\mathbf{e}}_0$ yields a smaller success probability. The complete proof has been detailed and commented in the following subsection (Sec.~\ref{ap:optimal}). Besides of showing the optimal success probability, these deductions affirm that the assistance of a third system is not required for the implementation of optimal parametric state separation and a two-dimensional ancilla suffices for this purpose.  Consequently, as expected, the use of a third system will not outperform unambiguous or maximum-confidence discrimination when carried out assisted by a two level ancilla.

\subsection{Proof of optimality of $p_\s(\xi)$ for parametrtic state separation \label{ap:optimal}}
\par Thousands of numerical simulations showed that the probability shown in Eq.~(\ref{eq:psxi}) is also the optimal probability that can be obtained by solving the optimization problem of Dunjko and Andersson [Eq.~(\ref{linear_problem})]. In order to show an analytical proof of this fact, let us study a solution in a neighborhood of the one of Eq.~(\ref{eq:psxi}), which can be written as ${\mathbf{x}_{0}=p_\s \vec{\mathbf{e}}_0}$ for the optimization algorithm since it was obtained from a leakless transform. Then, considering extra parameters $\kappa_0$ and $\epsilon_k$, the $\mathbf{x}$ unknown vector can be written as 
\begin{align}
	\mathbf{x} = (p_\s - \kappa_0) \vec{\mathbf{e}}_0 + \sum_{k=1}^{N-1} \epsilon_k \vec{\mathbf{e}}_k,\label{eq:Ansatz}
\end{align}
where the new unknowns are $\kappa_0$ and $\epsilon_k$ (for ${k=1,\dots,{N-1}}$). Bearing in mind, from Eq.~(\ref{eq:xvector}), that $\mathbf{x}$ has only non-negative coefficients, we must impose that $\kappa_0\leqslant p_\s$ and $\epsilon_{k}\geqslant 0$. Then, the linear optimization problem is now established as
\begin{align}
	\max_{\{\kappa_0,\epsilon_{k}\}}\hspace{0.7cm} &  p = p_\s - \kappa_0 + \sum_{k=1}^{N-1} \epsilon_{k}\nonumber \\
{\rm subject\,to}\hspace{0.7cm}& \Delta_{k} = a_k^2 - \sum_{s=0}^{N-1}b_{s}^2 x_{k-s}\geqslant 0,
\end{align}
where all vector indexes must be taken modulo $N$. By applying the proposal of Eq.~(\ref{eq:Ansatz}), we obtain 
\begin{align}
	\Delta_{k} = a_k^2 - (p_\s -\kappa_0)b_k^2 -\sum_{\substack{s=0\\s\neq k}}^{N-1}b_{s}^2 \epsilon_{k-s}\geqslant 0.\label{eq:better_cons}
\end{align}
\par The positivity of $\Delta_k$ from Eq.~(\ref{eq:better_cons}) must be fulfilled for every value of $k$. Particularly, when $k=k_{\min}$ such that ${a_{k_{\rm min}}=a_{\rm min}}$,  we obtain 
\begin{align}
	\kappa_0  & \geqslant \sum_{\substack{s=0\\s\neq k_{\min}}}^{N-1} \frac{b_{s}^2}{b_{\min}^2}\epsilon_{k_{\min}-s}  \geqslant \sum_{\substack{s=0\\s\neq k_{\min}}}^{N-1} y_{s}\epsilon_{k_{\min}-s} .\label{eq:kappapos}
\end{align}
For the case of linearly independent states, every coefficient $y_s$ is equal to one. Since $\epsilon_r\geqslant 0$, either {(i)}~${\kappa_0 = \epsilon_r = 0}$, or {(ii)}~$\kappa_0 >0$. The success probability for the first case is $p_{\rm (i)} = p_\s$, whereas for the second case is $p_{\rm (ii)} = p_\s -\kappa_0+\sum_r \epsilon_r$, considering $0 \neq \kappa_0\geqslant \sum_r \epsilon_r $. We can observe that
\begin{align*}
	p_{\rm (ii)} - p_{\rm (i)} =& -\kappa_0+\sum_r \epsilon_r \leqslant 0.
\end{align*}
Then, $p_{\rm (i)} \geqslant p_{\rm (ii)}$, showing that  $\kappa_0 = \epsilon_r = 0 $ is an optimal solution, which also satisfies Eq.~(\ref{eq:better_cons}) for $k\neq k_{\rm min}$. 
\par On the other hand, the case of linearly dependent states must be analyzed separately. Although for this case we have some coefficients $a_k$ equal to zero, Eq.~(\ref{eq:kappapos}) remains valid. Consequently, $\kappa_0$ must be a non-negative number. Now, let $\mathcal{Q}=\{\mu =0,\dots,N-1|y_\mu=0\}$ be the set of indexes related to vanishing coefficients of the fiducial states. The optimal probability is  
\begin{align}
p &= p_\s- \kappa_0+\sum_{\substack{s\not\in\mathcal{Q}\\s\neq k_{\min}}} \epsilon_{k_{\min}-s} + \sum_{\mu \in\mathcal{Q}} \epsilon_{k_{\min}-\mu} \nonumber\\
 &\leqslant p_\s + \sum_{\mu \in\mathcal{Q}} \epsilon_{k_{\min}-\mu},\label{eq:popt_ld}
\end{align}
where the inequality of Eq.~(\ref{eq:kappapos}) was applied. Let ${Q=N-D}$ be the number of elements of $\mathcal{Q}$. As it transpires from Eq.~(\ref{eq:popt_ld}), the bounds on the optimal probability now depend on $Q$ parameters $\epsilon_r$. For $\mu\in\mathcal{Q}$, Eq.~(\ref{eq:better_cons}) becomes
\begin{align}
	\Delta_{\mu} &= -\sum_{\substack{s=0\\s\neq q}}^{N-1} b_{s}^2 \epsilon_{\mu-s} = -\sum_{\substack{s\not\in\mathcal{Q}}} b_{s}^2 \epsilon_{\mu-s} \geqslant 0,\label{eq:deltaq}
\end{align}
which can be satisfied only if $\forall\,s\not\in\mathcal{Q},\,\,\epsilon_{\mu-s}=0$. In other words, $y_s\neq 0$ implies that $\epsilon_{\mu-s}=0$, setting the value of $D$ of the total of ${N-1}$ available $\epsilon_r$ as zero. When $N=D+1$, this condition imposes all $\epsilon_r$ must be zero, preventing the optimal probability of Eq.~(\ref{eq:popt_ld}) from being greater than $p_\s$. Similarly, for $Q=2$ we have $\mathcal{Q}=\{\mu_1,\mu_2 \}$. Thus, $\Delta_{\mu_1}\geqslant 0$ implies that only $\epsilon_{\mu_1 -\mu_2}$ \emph{might} differ from zero and $\Delta_{\mu_2}\geqslant 0$ indicates that only $\epsilon_{\mu_2 -\mu_1}$ \emph{might} not vanish. Together, these two statements suggest that every $\epsilon_r$, including $\epsilon_{\mu_2 -\mu_1}$ and $\epsilon_{\mu_1 -\mu_2}$, are zero \emph{unless} $\mu_1 -\mu_2 \equiv \mu_2 -\mu_1\,{\rm mod\,}N$, for which no information is given. Although interesting, this last case does not exert influence on the upper bound of Eq.~(\ref{eq:popt_ld}) since it contains only terms of the type $\epsilon_{k_{\min}-\mu}$. Reminding that $k_{\min}\not\in\mathcal{Q}$, these terms in the upper bound vanish as consequence of $\Delta_{\mu_1}\geqslant 0$ and $\Delta_{\mu_2}\geqslant 0$, as aforementioned. Then, the optimal probability cannot surpass $p_\s$. In the same way, Eq.~(\ref{eq:deltaq}) for $Q > 2$ constructs a set of $Q$ inequalities, each one imposing $D$ of the $\epsilon_r$ to adopt null value. For every $\mu_j \in\mathcal{Q}$, $\Delta_{\mu_j}\geqslant 0$ informs that only coefficients of the class $\epsilon_{\mu_j -\mu_k}$ \emph{might} adopt values different from zero, where $\mu_k$ are other elements of $\mathcal{Q}$. Otherwise, any other $\epsilon_r$ must be zero and the upper bound of Eq.~(\ref{eq:popt_ld}) cannot exceed  $p_\s$ since terms like $\epsilon_{k_{\min}-\mu}$ does not belong to the type of coefficients $\epsilon_{\mu_j -\mu_k}$ that could have been greater than zero. For all of these studied cases, the fact that the optimal probability is $p_\s$ means that ${\kappa_0=\epsilon_r=0}$. This result that also satisfies the constraints of Eq.~(\ref{eq:better_cons}) when ${k\neq k_{\min}}$ and ${k\not\in\mathcal{Q}}$. This completes the proof. \hfill{$\blacksquare$}\\

\par We have proven that $\mathbf{x} = p_\s \vec{\bf e}_0$ is the optimal solution of the optimization problem for this class of state separation, for both linearly independent and linearly dependent states, since any deviation $\sum_{r}\epsilon_r\vec{\bf e}_r -\kappa_0\vec{\bf e}_0$ from the solution $p_\s\vec{\bf e}_0$ yields a smaller probability. In principle, this solution is a local maximum. Nevertheless, the local extrema of linear functions are also global extrema. The solution found is, then, the global maximum of the probability of success.

\par It is noteworthy that the result from this optimization does not depend on the explicit parametrization of Eq.~(\ref{eq:bkcoef}) possibly as a consequence of the preserved ordering of the coefficients. As an additional comment, these results also show that the use of a three-system strategy for unambiguous discrimination of symmetric states does not overcome the use of only two systems. The same conclusion holds for maximum confidence discrimination among symmetric states.

\section{Application to Quantum Teleportation and Entanglement Swapping \label{SS_applications}}

\par In quantum teleportation, an unknown qubit state is deterministically and faithfully transmitted from a sender (Alice) to a receiver (Bob) with the consumption of a shared maximally entangled two-qubit state (the quantum channel) and two bits of forward classical communication~\cite{Bennet93}. Otherwise, when the quantum channel is not maximally entangled, faithful teleportation is achieved only probabilistically, whereas a deterministic implementation implies in less than unit average fidelity for the protocol. In either case one seeks for optimized schemes. On one hand perfect conclusive teleportation~\cite{Mor96,Gu02} achieves unit fidelity with maximum success probability. On the other hand, with the standard protocol~\cite{Bennet93} one achieves the optimal average teleportation fidelity deterministically~\cite{Banaszek00,Vidal00BipTr}. 
\par Quantum teleportation of pure qudit states ($D$-level quantum systems) using non-maximally entangled channels comprises the problem of discriminating among a set of equally likely symmetric states, whose linear dependence is based on the number of non-vanishing coefficients of the entangled state used as a channel. According to this, unambiguous discrimination or maximum confidence discrimination can be applied to teleport, probabilistically, with the maximum achievable fidelity~\cite{Roa03,Neves12,SolisProsser13}. Otherwise, the standard procedure is equivalent to perform minimum error discrimination and the teleportation is accomplished deterministically with an average fidelity lower than the probabilistic case~\cite{Ban97,Barnett01sym}. In this regard, we can resort to state separation of symmetric states to interpolate between these two cases, i.e., increase the success probability compared to the former and the teleportation fidelity compared to the latter. Briefly, consider a two-qudit state given by 
\begin{equation}
|\Psi\rangle_{12}=\sum_{m=0}^{N-1}a_m|m\rangle_{1}|m\rangle_{2},
\label{QUANTUMCHANNEL}	
\end{equation}
where $a_m$ are Schmidt coefficients, non-negative real numbers satisfying  $\sum_{m=0}^{N-1}a_m^2=1$. The number of non-vanishing Schmidt coefficients is $D$ ($\leqslant N$). Let $|\phi\rangle_3$ be a $N$-dimensional state to be teleported. Particle 1 is in Bob's possession, whereas Alice owns particles 2 and 3. Quantum teleportation can be described by, firstly, Alice applying a $\hat{G}^\text{\sc xor}_{23}$ operation on the pair of particles in her possession. The state of the three-particle system becomes  
\begin{equation}
\hat{G}^\text{\sc xor}_{23}|\Psi\rangle_{12}|\phi\rangle_3=
\frac{1}{N}\sum_{l,k=0}^{N-1}\hat{Z}_1^{N-l}\hat{X}_1^{k}|\phi\rangle_1|\alpha_l\rangle_2|k\rangle_3,
\label{BASE}
\end{equation}
where $\hat{G}^\text{\sc xor}_{23}$ is a generalization of a CNOT gate, whose action on the computational basis is defined by $\hat{G}^\text{\sc xor}_{23}|i\rangle_2|j\rangle_3=|i\rangle_2|i\ominus j\rangle_3$, where the addition $\oplus$ and subtraction $\ominus$ are modulo $N$ operations~\cite{Alber01,Jex03}. The symmetric states $|\alpha_j\rangle$ are given by Eq.~(\ref{eq:alphaj}) with $\phi_k=0$, whose fiducial coefficients are the Schmidt coefficients of the entangled state used as channel. The operators $\hat{Z}$ and $\hat{X}$ are the generalized Pauli operators described in Section~\ref{sec:intro_symm}. The standard procedure consists now in Alice performing von Neumann measurements on particles 2 and 3 in the inverse Fourier and computational bases, respectively. Alice, then, communicates her results to Bob in order for him to apply necessary unitary operations to obtain the state. If, instead, Alice decides to separate the states $|\alpha_{j}\rangle$ before measuring, she will obtain, from Eqs.~(\ref{eq:unitary}) and~(\ref{BASE}), 
\begin{widetext}
\begin{align}
\mathcal{U}_{2a}(\xi)\hat{G}^\text{\sc xor}_{23}|\Psi\rangle_{12}|\phi\rangle_3 |\mathscr{A}\rangle_a =&
\frac{\sqrt{p_\s(\xi)}}{N}\sum_{l,k=0}^{N-1}\hat{Z}_1^{N-l}\hat{X}_1^{k}|\phi\rangle_1|\beta_l(\xi)\rangle_2|k\rangle_3|0\rangle_a + \frac{\sqrt{1-p_\s(\xi)}}{N}\sum_{l,k=0}^{N-1}\hat{Z}_1^{N-l}\hat{X}_1^{k}|\phi\rangle_1|\widetilde{\beta}_l\rangle_2|k\rangle_3|1\rangle_a.
\label{BASE2}
\end{align}
\end{widetext}
Therefore, with success probability $p_\s(\xi)$ [Eq.~(\ref{eq:psxi})], Alice can teleport the state as they had a bipartite entangled state with Schmidt coefficients $b_k(\xi)$---which are present in states $|\beta_{j}(\xi)\rangle$---instead of the original $a_k$. The probabilistic average fidelity is~\cite{Horodecki99,Vidal00BipTr,Banaszek00,Banaszek01} 
\begin{align}
	\mathscr{F}_{\rm ave}(\xi)=\frac{1}{1+N}\left[1+\left(\displaystyle\sum_{k=0}^{N-1}b_k(\xi)\right)^2\right].
\end{align}
We must note that $\xi=0$ recovers the standard procedure and $\xi=1$ stands for using unambiguous or maximum-confidence discrimination~\cite{Neves12,SolisProsser13}. It can be seen that the number of Schmidt coefficients different from zero is meaningful for $\mathscr{F}_{\rm ave}(\xi)$, since it can attain ${(D+1)/(N+1)}$ as its maximum value.

\par
\begin{figure}
\centering
\includegraphics[height=6.1cm]{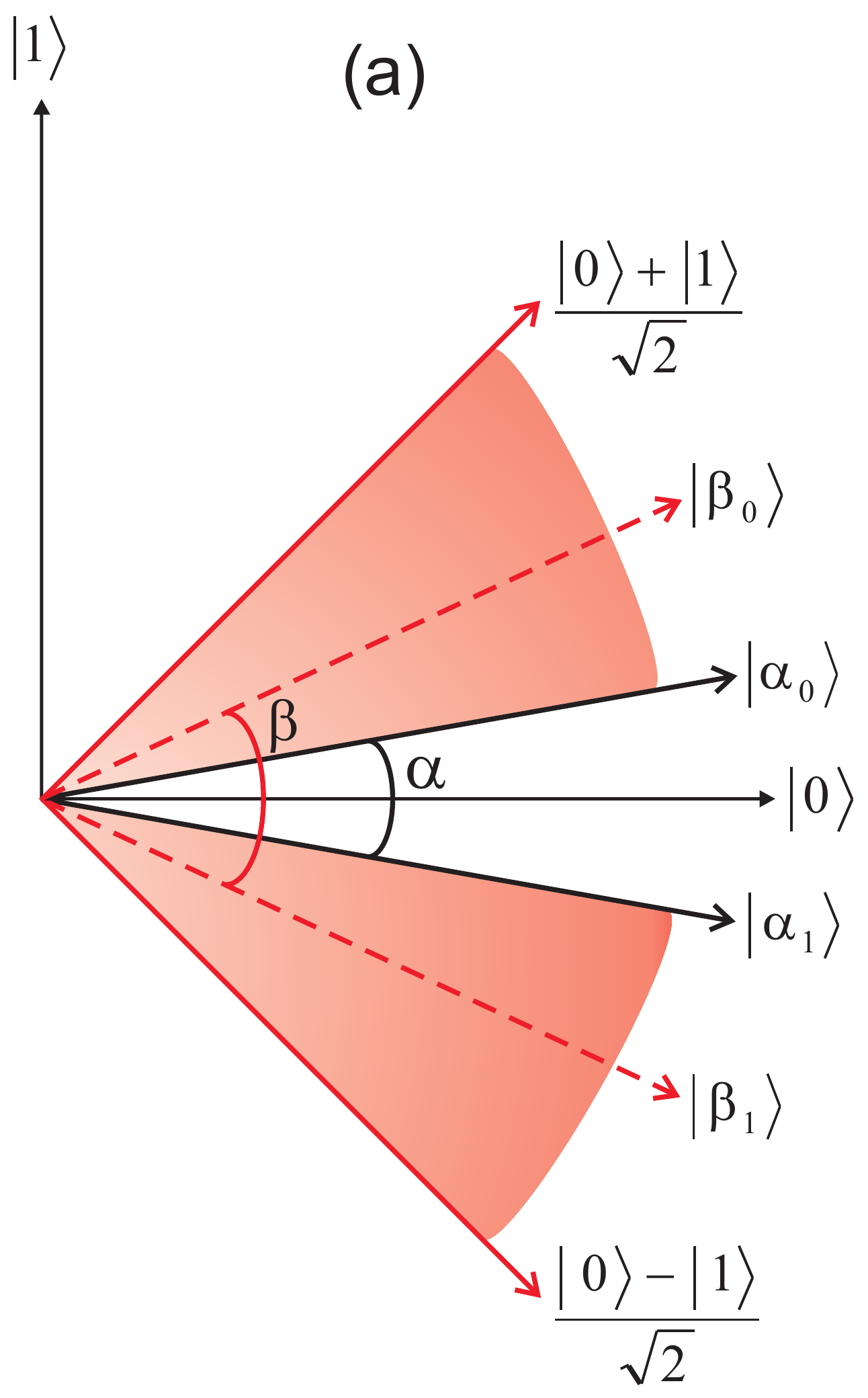}
\includegraphics[height=6.1cm]{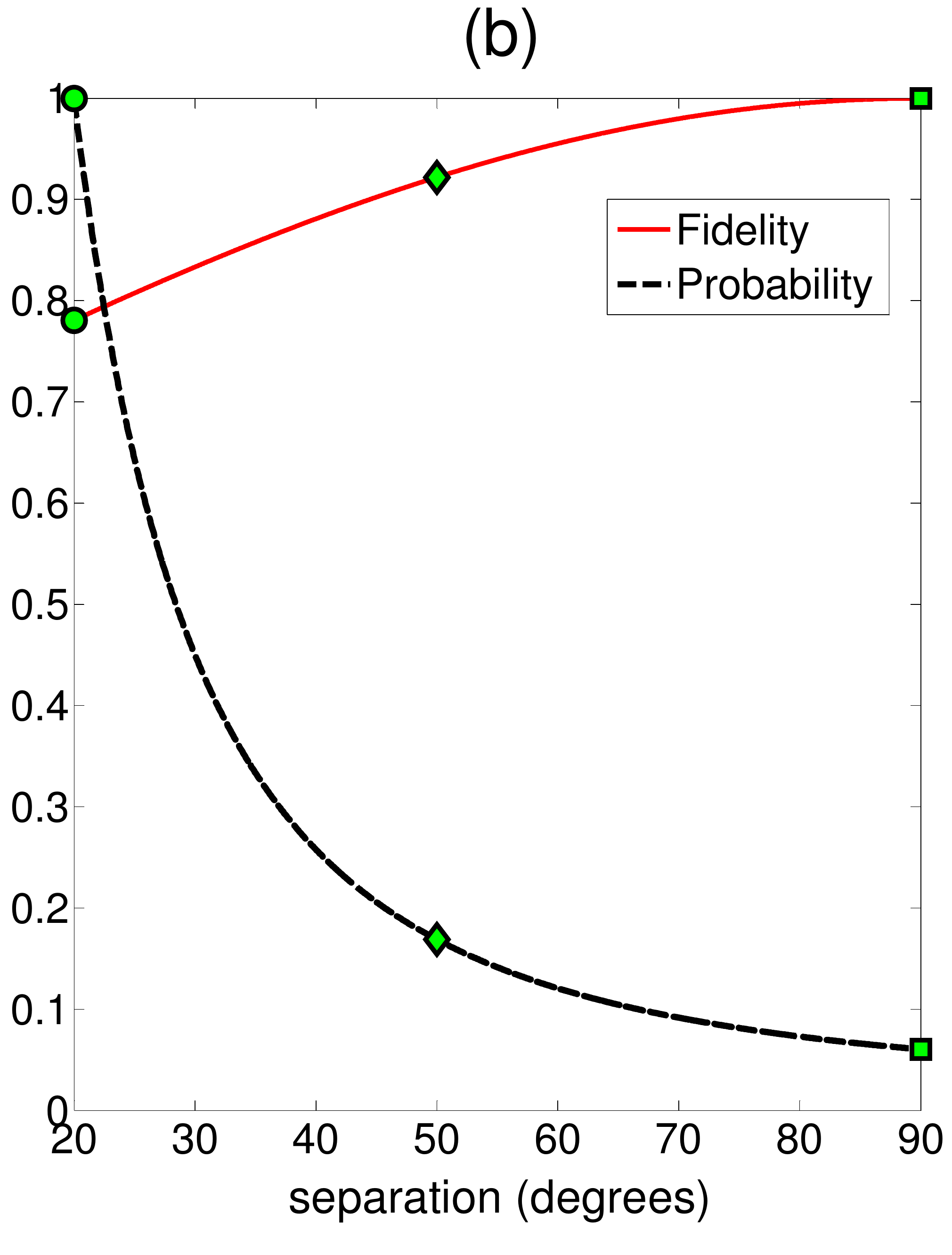}
\caption{(Color online) (a) The initial separation $\alpha$ can be probabilistically increased to any angle $\beta$ within the shaded region. (b) For $\alpha = 20^\circ$ the curves show the success probability and average teleportation fidelity as a function of the separation angle $\beta$. The symbols indicate the
following protocols: (circles) optimal deterministic, (squares) perfect conclusive and (diamonds) imperfect conclusive for $\beta= 50^\circ$. \label{fig:alpha01}}
\end{figure}

 As an illustrative example, we can review this procedure applied to ${N=D=2}$. A useful decomposition for $|\alpha_j\rangle$ ($j=0,1$) is $|\alpha_j\rangle=\cos(\alpha/2)|0\rangle+e^{i\pi j}\sin(\alpha/2)|1\rangle$, with $0\leqslant \alpha \leqslant \pi/2$, so that $\alpha=\cos^{-1}\langle\alpha_0|\alpha_1\rangle$ is the angle between $|\alpha_0\rangle$ and $|\alpha_1\rangle$, as Fig.~\ref{fig:alpha01}(a) shows. The standard teleportation procedure, also known as optimal deterministic teleportation, can teleport states with average fidelity $\mathscr{F}_{\rm ave}^{(\alpha)}=[2+\sin(\alpha)]/3$~\cite{Vidal00BipTr,Banaszek00}. In order to attain a fidelity equal to one, in a process known as perfect conclusive teleportation~\cite{Mor96,Gu02}, Alice discriminates the states $|\alpha_j\rangle$ unambiguously with the optimal success probability~\cite{Ivanovic87,Dieks88,Peres88} 
\begin{align}
	P_\alpha = 1-\langle\alpha_0|\alpha_1\rangle=2\sin^2(\alpha/2),\label{eq:p_alpha}
\end{align}
which is then the probability of a successful teleportation. She tells Bob whether the discrimination succeeded or not by sending him one classical bit. When it succeeds, the protocol is accomplished as in the standard case and the teleported state has unit fidelity. Otherwise, in case of failure they interrupt the process. 
\par Now, we will analyze a third approach that can be recognized as \emph{imperfect conclusive teleportation}, in which an imperfect replica of state $|\phi\rangle$ is obtained after teleportation assisted by a conclusive attempt of state separation. Let $|\beta_0\rangle$  and $|\beta_1\rangle$ be two states separated by angle $\beta$, as illustrated in Fig.~\ref{fig:alpha01}(a). In this process, the states $|\alpha_j\rangle$ are probabilistically mapped onto $|\beta_j\rangle$ through parametric state separation, with $\xi=1-\cos(\beta)/\cos(\alpha),$ and a success probability of~\cite{Chefles98qss}
\begin{align}
	P_\beta= \frac{1-\langle\alpha_0|\alpha_1\rangle}{1-\langle\beta_0|\beta_1\rangle} = \frac{P_\alpha}{2\sin^ 2(\beta/2)},
\end{align}
which, as expected, coincides with the optimal probability of Eq.~(\ref{eq:psxi}) with $a_{\min}=\sin(\alpha/2)$ and  $\xi=1-\cos(\beta)/\cos(\alpha).$ This is the probability that the teleportation will be carried out. It is easy to see that $P_\beta\geqslant P_\alpha$. Alice sends Bob one classical bit communicating whether the
mapping $|\alpha_j\rangle\rightarrow |\beta_j\rangle$ succeeded or not. When it fails, they interrupt the protocol; otherwise, the protocol proceeds as in the standard case. The average teleportation fidelity is~\cite{Vidal00BipTr,Banaszek00}
\begin{align}
	\mathscr{F}_{\rm ave}^{(\beta)}=[2+\sin(\beta)]/3,
\end{align}
which is the optimal one and satisfies $\mathscr{F}_{\rm ave}^{(\beta)}\geqslant\mathscr{F}_{\rm ave}^{(\alpha)}$. 

Now, let us assume that for Alice and Bob purposes it is sufficient to accomplish teleportation with a fidelity of transmission better than a given threshold. In addition, they have scarce resources (e.g. entangled states) and share a quantum channel with a small degree of entanglement (which means small $\alpha$). In this scenario, standard teleportation may not be enough to overcome that threshold deterministically while perfect conclusive teleportation would not be a clever choice for them since unit fidelity is not required and the probability of success, $P_\alpha$, is small. The best thing to do is for Alice to apply state separation on qubit 2 by setting a separation angle $\beta~(< \pi/2)$ that, when successful, is just enough to enable an average teleportation fidelity above the threshold. Doing so, they will increase their transmission efficiency. This situation can be illustrated from Fig.~\ref{fig:alpha01}(b). Assume a fidelity threshold $\mathscr{F}_{\rm thr} = 0.91$. For the quantum channel considered there with $\alpha = 20^\circ$, the deterministic protocol delivers a fidelity $\mathscr{F}_\alpha \approx 0.78$ (circle), while the perfect conclusive one succeeds with probability $P_\alpha \approx 0.06$ (square). In this case, an imperfect conclusive protocol that gives an average teleportation fidelity $\mathscr{F}_\beta \approx 0.92$, will succeed with probability $P_\beta \approx 0.17$ (diamonds).

\par An analogous analysis can be scrutinized for entanglement swapping. Unlike the previous case, the description of this protocol is more complicated and, for the sake of simplicity and ease of comprehension, only basic details are given. In entanglement swapping, two users have, each one, a two-qudit entangled state, say $|\kappa\rangle_{12}$ for Alice and $|\chi\rangle_{34}$ for Bob. Each party sends one particle, say 1 and 4, to a third party who will perform a Bell state measurement on them, preparing an entangled state shared along two parties that never interacted, i.e., an entangled state $|\tau\rangle_{23}$~\cite{Bennet93,Zukowski93}. When at least one of the initial bipartite states ($|\kappa\rangle_{12}$ or $|\chi\rangle_{34}$) is not maximally entangled, the resulting state $|\tau\rangle_{23}$ will not be maximally entangled either. A deterministic implementation is related with an optimal minimum-error strategy, while unambiguous (or maximum-confidence) discrimination enables one to accomplish the protocol with the maximum achievable entanglement for the quantum channels involved~\cite{Delgado05,SolisProsser14}. With the state separation presented here one can interpolate between these two extreme cases, trading off success probability and the amount of entanglement of the states resulting from swapping.

\section{Proposal for optical implementation \label{SSsec:impl1}}

\begin{figure}[!t]
	\centering
	\includegraphics[width=0.45\textwidth]{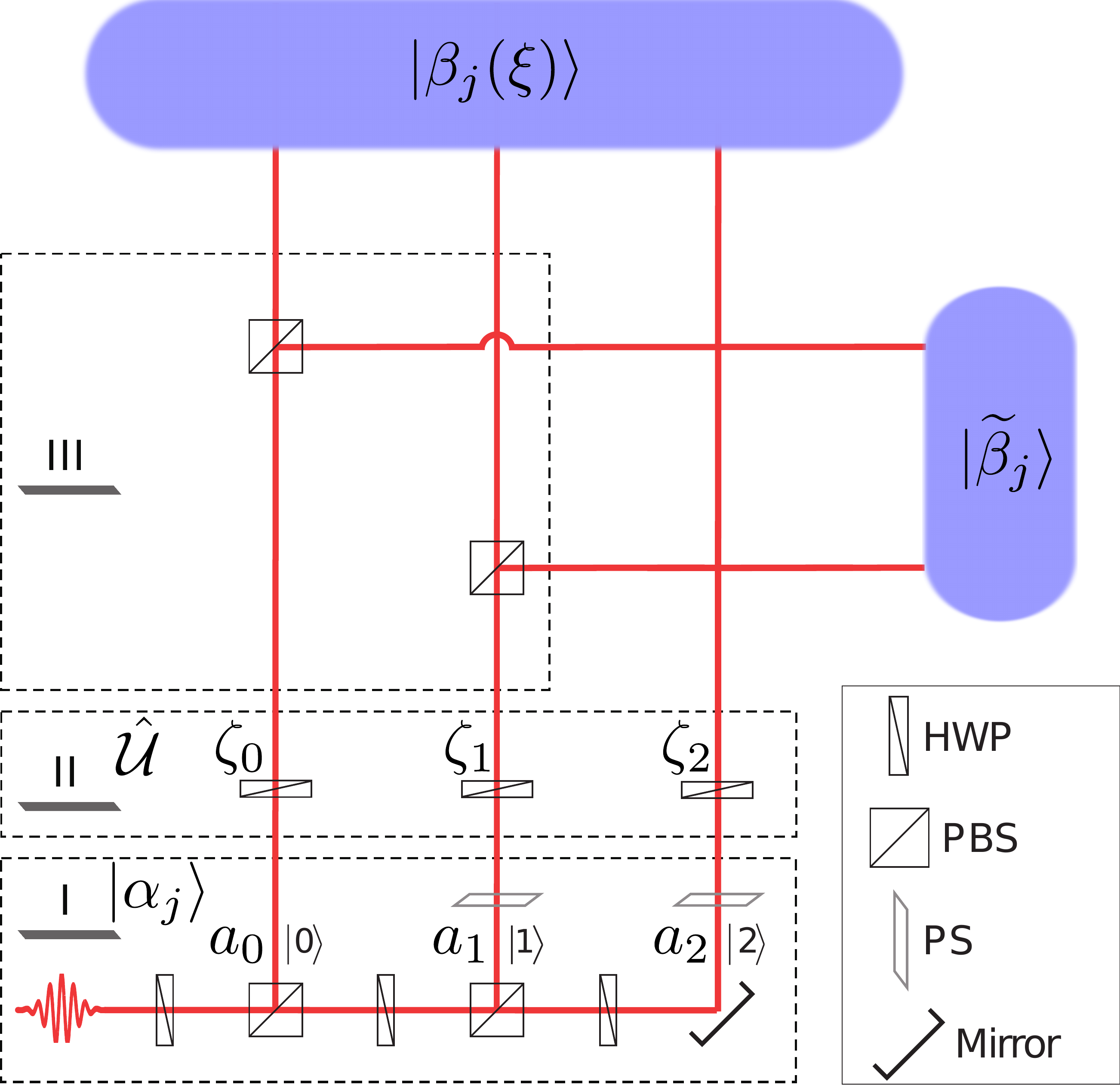}
	\caption{(Color online) Experimental proposal for implementation of quantum state separation of $N$ states from a three-dimensional Hilbert space. HWP: half-wave plate. PBS: polarizing beam splitter. PS: phase shifter (see text for details).\label{drawing}}
\end{figure}

\par Figure~\ref{drawing} illustrates a simplified optical setup able to perform optimal parametric state separation. This particular case considers $\phi_k=\varphi_k=0,~~\forall~k$ [see Eqs.~(\ref{eq:alphaj}) and~(\ref{eq:betasuc})].  Half-wave plates (HWPs) are used to assist the state preparation and to implement conditional operations depending on which path the photon followed. A HWP whose fast axis is oriented at an angle $\zeta$ with respect to the horizontal axis is represented by a Jones matrix given by~\cite{SalehBook}
\begin{align}
	\hat{H}(\zeta)=\begin{pmatrix}
		\cos(2\zeta)	& \sin(2\zeta) \\ \sin(2\zeta) & -\cos(2\zeta)
	\end{pmatrix},
\end{align}
written in the $\{|h\rangle,|v\rangle\}$ basis. Stage I is aimed to prepare the symmetric states encoded in the longitudinal spatial modes of a single photon, whereas its polarization is the degree of freedom used as ancilla. Phase shifters are used to encode the phases corresponding to each symmetric state. The state of the photon after this stage is $|\alpha_{j}\rangle|v\rangle$, considering that polarizing beam splitters (PBSs) reflect incident vertically polarized photons and transmit horizontally polarized photons. In Stage II, the unitary operation of Eq.~(\ref{eq:unitary}) is implemented by the HWP at each path. This operation can be written as
\begin{align}
	\hat{\mathcal{U}} =& \sum_{k=0}^{D-1}|k\rangle\langle k|\otimes \hat{H}(\zeta_k)\nonumber \\
	=& \left(\sum_{k=0}^{D-1}\cos(2\zeta_k)|k\rangle\langle k|\right)\otimes\left(|h\rangle\langle h|-|v\rangle\langle v|\right) \nonumber \\ &+ \left(\sum_{k=0}^{D-1}\sin(2\zeta_k)|k\rangle\langle k|\right)\otimes\left(|h\rangle\langle v|+|v\rangle\langle h|\right).
\end{align}

By setting the angles of each HWP such that
\begin{align}
	\sin(2\zeta_k) =& ~~~~\sqrt{ \frac{1-\xi+\xi/{ D }a_k^2}{1-\xi+\xi/{ D }a_{\rm min}^2} } ,\\
	\cos(2\zeta_k) =& -\sqrt{ \frac{\xi}{{ D }} \frac{ 1/a_{\rm min}^2 - 1/a_k^2}{1-\xi+\xi/{ D }a_{\rm min}^2} },
\end{align}
we obtain
\begin{align}
	\hat{\mathcal{U}} =& \hat{A}_\s\otimes\left(|h\rangle\langle v|+|v\rangle\langle h|\right) + \hat{A}_\f\otimes\left(|v\rangle\langle v|-|h\rangle\langle h|\right),
\end{align}
which reduces to Eq.~(\ref{eq:unitary}), considering ${|\mathscr{A}\rangle_a=|v\rangle}$, ${|0\rangle_a=|h\rangle}$, and ${|1\rangle_a=|v\rangle}$, when applied on ${|\alpha_{j}\rangle|v\rangle}$. In this case, failure states $|\widetilde{\beta}_{j}\rangle$ [Eq.~(\ref{eq:betafail})] do not depend on $\xi$ since null phases for the fiducial states were chosen. Finally, Stage III performs a measurement on photon polarization which, in case of a successful event, yields a photon prepared in state $|\beta_j(\xi)\rangle$, shown above Stage III. Otherwise, the state will be $|\widetilde{\beta}_j\rangle$ as sketched at the right side of Stage III.

\section{Conclusions \label{SSsummary}}
\par Besides minimum-error, unambiguous discrimination and maximum-confidence as strategies for discrimination of non-orthogonal states, there exist other strategies that interpolate between minimum-error and unambiguous discrimination for linearly independent states, and between minimum-error and maximum-confidence for linearly dependent ones. These strategies usually impose either a fixed value or an upper bound for the probability of mistaking the retrodiction of the state~\cite{Touzel07,Hayashi08,Sugimoto09,Sugimoto12,Sentis13} or for the probability of a failed attempt~\cite{Chefles98fixed,Bagan12,Nakahira12,Herzog12} while optimizing the probability of conclusive and correct results, respectively. Although these strategies comprise different approaches for quantum state discrimination and different optimization problems, it has been shown that they are closely related and the optimization of the probability of correct results under a fixed rate of inconclusive results can be found by optimizing the same probability of correct results under a certain fixed error probability and vice versa~\cite{Herzog12}. Moreover, it has been shown~\cite{Nakahira12} that a given set of measurements aimed to discriminate quantum states with a particular failure probability attains the optimal probability of correct results if and only if the whole discrimination process can be described by a map from the original set of states onto a specific set of new states and then performing optimal minimum-error probability on the latter (Theorem~1 of Ref.~\cite{Nakahira12}). This finding allows a new approach to the aforementioned optimization problems by treating them all as maps between sets of quantum states followed by minimum-error discrimination. Concerning to the topic studied throughout this paper, we can use the probability of success [Eq.~(\ref{eq:psxi})] to set a fixed rate of inconclusive results and find the value of $\xi$, which can be labeled as $\xi_{\rm fixed}$, that fits with this constraint. Afterwards, the use of Eqs.~(\ref{eq:As}) and~(\ref{eq:Af}) allow to find the Kraus operators needed to map from $|\alpha_j\rangle$ to $|\beta_j(\xi_{\rm fixed})\rangle$. Since these new states are also equally likely symmetric states, the solution for optimal minimum-error discrimination among them is known in the literature~\cite{Ban97,Barnett01sym}. Thus, discrimination strategies interpolating between minimum-error and unambiguous (or maximum-confidence) discrimination can benefit from the map proposed in this work. It is worth mentioning that the states resulting from a failed attempt, which are given by Eq.~(\ref{eq:betatilde}), are also symmetric pure states. This opens the possibility of studying sequential maps analogous to the sequential discrimination proposed in Ref.~\cite{Jimenez11} in which it is still possible to use the inconclusive results to attempt discrimination instead of just discarding these results. 

\par Summarizing, we have proposed a map between sets of uniparametric symmetric pure quantum states that leads to more distinguishable states. The results reported here consider both linearly independent and dependent states. The optimal probability of success was obtained and applications to teleportation of quantum states and entanglement swapping were shown. 
	
\par Likewise teleportation, entanglement swapping with non-maximally entangled states can be assisted by state separation of symmetric states. Thus several applications of entanglement swapping can also be improved with the protocol presented here, as for instance, quantum repeaters~\cite{Sangouard11}, entanglement generation between distant users~\cite{Briegel98,Dur99,Waks02}, some entanglement concentration schemes~\cite{Shi00,Hsu02,Modlawska08,Yang09}, experimental studies on nonlocality~\cite{Zukowski93}, generation of Greenberger-Horne-Zeilinger (GHZ) states via multiparticle entanglement swapping~\cite{Hardy00,Bose98}, quantum secret sharing~\cite{Zhou09}, and other quantum communication protocols~\cite{Xia07,Zhan09,Qin10,Scherer11,Zhou11}.

\begin{acknowledgments}
This work was supported by CONICyT Grant No. PFB-0824, Millenium Scientific Initiative Grant No. RC130001, and FONDECYT grants 11085057 and 1140635. M.A.S.P. acknowledges the financial support from CONICYT. L. N. acknowledges financial support from Brazilian agencies CNPq (grant 485401/2013-4), FAPEMIG (grant APQ-00149-13) and PRPq/UFMG (grant ADRC-01/2013). O.J. acknowledges financial support from FONDECYT grant 11121318. 
\end{acknowledgments}

\appendix*
\section{Optimization and Kraus operators \label{ap:kraus}}

\par The linear programming problem of Eq.~(\ref{linear_problem}) offers a solution to the task of mapping one set of symmetric states onto another. Although the solution is explicit in terms of probability and the leak of the transform, no information about the Kraus operators is provided. Equation~(\ref{eq:three_system}) shows the Stinespring representation of the linear map, although determining a Kraus operator decomposition is desirable as well. So, let us consider a set of $N+1$ Kraus operators $\{\hat{t}_{\f},\hat{t}_{k \s}\}_{k=0}^{N-1}$ such that~\cite{KrausChapter} 
\begin{equation}
\sum_{k=0}^{N-1}(\hat{t}_{k \s}^\dagger \hat{t}_{k \s}^{\vphantom{\dagger}})+\hat{t}_{\f}^\dagger \hat{t}_{\f}^{\vphantom{\dagger}}=\hat{I},
\end{equation}
where $\hat{t}_{k \s}$ and $\hat{t}_{\f}$ stand for the successful and unsuccessful cases, respectively. The matrix elements of each operator $\hat{t}_{k \s}$ can be calculated as~\cite{Keyl02}
\begin{align}
	\langle y|\hat{t}_{k\s}|z\rangle = \langle y\otimes k\otimes 0|\mathcal{\hat{U}}|z \otimes \mathcal{M}\otimes \mathcal{A}\rangle.
\end{align}
From Eq.~(\ref{eq:three_system}) we have that 
\begin{align*}
\mathcal{\hat{U}}\sum_{k'=0}^{N-1}a_{k'} &\omega^{j k'}|k'\rangle_s|\mathcal{M}\rangle_e|\mathcal{A}\rangle_a \\  =& \sqrt{p}\sum_{l,m=0}^{N-1}b_{l} \Psi_{m} \omega^{j(l+m)}|l\rangle_s|m\rangle_e|0\rangle_a \\&\hspace{2cm}+ \sqrt{1-p}|\Phi_{j}\rangle_{se}|1\rangle_a. 
\end{align*}
Applying $\langle y\otimes k\otimes 0 |$ from the left, we obtain 

\begin{align*}
\sum_{k'=0}^{N-1} &a_{k'}\omega^{k'j} \langle y\otimes k\otimes 0 |\mathcal{\hat{U}}|k' \otimes \mathcal{M}\otimes \mathcal{A}\rangle \\=& \sqrt{p}\sum_{l,m=0}^{N-1}b_{l} \Psi_{m} \omega^{j(l+m)}\langle y\otimes k\otimes 0|l\otimes m\otimes 0\rangle \\ &\hspace{3cm} + \sqrt{1-p}\langle y\otimes k\otimes 0 |\Phi_{j}\rangle_{se}|1\rangle_a , 
\end{align*}
or, equivalently,
\begin{align*}
\sum_{k'=0}^{N-1}a_{k'} \omega^{k'j}  \langle y|\hat{t}_{k\s}|k'\rangle =& \sqrt{p}b_{y} \Psi_{k} \omega^{j(y+k)} .
\end{align*}
The $\langle y|\hat{t}_{k\s}|z\rangle$ components can be isolated by applying inverse Fourier transform ($\sum_{j=0}^{N-1}\omega^{-jz}/\sqrt{N}$). This leads us to obtain 
\begin{align*}
 \langle y|\hat{t}_{k\s}|z\rangle =& \sqrt{p} \frac{b_y \Psi_k}{a_{z}} \delta_{y+k,z}.
\end{align*}
Thus, 
\begin{align*}
 \hat{t}_{k\s} 	=& \sum_{y,z=0}^{N-1}|y\rangle\langle z| \langle y|\hat{t}_{k\s}|z\rangle = \sqrt{p} \sum_{y=0}^{N-1} \frac{b_y \Psi_k}{a_{y+k}}  |y\rangle\langle y+k|.
\end{align*}
The Kraus operators $\hat{t}_\f$ can be obtained, although not uniquely, by 
\begin{align}
\hat{t}_{\f} =& \left[ \hat{I} - \sum_{y=0}^{N-1} \hat{t}_{k\s}^\dagger \hat{t}_{k\s}^{\vphantom{\dagger}} \right]^{1/2},
\end{align}
and, recalling that $\mathbf{x}=p\sum_{k=0}^{N-1}\Psi_k^2|k\rangle$, we then have
\begin{align}
\hat{t}_{k\s} =& \sqrt{x_k} \sum_{y=0}^{N-1} \frac{b_y}{a_{y+k}}  |y\rangle\langle y+k|.
\end{align}
In consequence, the results obtained from the linear optimization program reveals not only the optimal probability but also the number of Kraus operators required for the successful case, which is the number of non-vanishing components of $\mathbf{x}$. So, the $|\psi_{y}\rangle$ vectors contain information about the physical process in terms of Kraus operators. As a safety check, it can be readily shown that 
\begin{align*}
\sum_{k=0}^{N-1}\hat{t}_{k\s}^{\vphantom{\dagger}} |\alpha_j\rangle\langle\alpha_j|\hat{t}_{k\s}^\dagger = p |\beta_j\rangle\langle\beta_j|.
\end{align*}
We have considered only one Kraus operator $\hat{t}_\f$ for the inconclusive case. Although there can be several failure Kraus operators $\hat{t}_{r\f}$, we have considered that ${\hat{t}_{\f}^\dagger \hat{t}_{\f}^{\vphantom{\dagger}} = \sum_{r} \hat{t}_{r\f}^\dagger \hat{t}_{r\f}^{\vphantom{\dagger}} }$ since the main interest remains focused on the Kraus operators associated with successful attempts.

\end{document}